\documentclass[11pt]{article}

\usepackage{amssymb}
\usepackage{amsfonts}
\usepackage[arrow,matrix]{xy}

\makeatletter
\def\flE{\begin{picture}(0,0)
   \put( 0.25,    0){\vector( 1, 0){0.50}}
   \@ifstar{\@flE}{\@@flE}}
\def\@flE  #1{\put( 0.5 ,-0.03){\makebox(0,0)[ t]{$#1$}}\end{picture}}
\def\@@flE #1{\put( 0.5 , 0.03){\makebox(0,0)[ b]{$#1$}}\end{picture}}
\def\flNE{\begin{picture}(0,0)
   \put( 0.18, 0.18){\vector( 1, 1){0.64}}
   \@ifstar{\@flNE}{\@@flNE}}
\def\@flNE #1{\put( 0.52, 0.48){\makebox(0,0)[tl]{$#1$}}\end{picture}}
\def\@@flNE#1{\put( 0.48, 0.52){\makebox(0,0)[br]{$#1$}}\end{picture}}
\def\flN{\begin{picture}(0,0)
   \put(    0, 0.20){\vector( 0, 1){0.60}}
   \@ifstar{\@flN}{\@@flN}}
\def\@flN  #1{\put( 0.03, 0.5 ){\makebox(0,0)[ l]{$#1$}}\end{picture}}
\def\@@flN #1{\put(-0.03, 0.5 ){\makebox(0,0)[ r]{$#1$}}\end{picture}}
\def\flNW{\begin{picture}(0,0)
   \put(-0.18, 0.18){\vector(-1, 1){0.64}}
   \@ifstar{\@flNW}{\@@flNW}}
\def\@flNW #1{\put(-0.48, 0.52){\makebox(0,0)[bl]{$#1$}}\end{picture}}
\def\@@flNW#1{\put(-0.52, 0.48){\makebox(0,0)[tr]{$#1$}}\end{picture}}
\def\flW{\begin{picture}(0,0)
   \put(-0.25,    0){\vector(-1, 0){0.50}}
   \@ifstar{\@flW}{\@@flW}}
\def\@flW  #1{\put(-0.5 , 0.03){\makebox(0,0)[ b]{$#1$}}\end{picture}}
\def\@@flW #1{\put(-0.5 ,-0.03){\makebox(0,0)[ t]{$#1$}}\end{picture}}
\def\flSW{\begin{picture}(0,0)
   \put(-0.18,-0.18){\vector(-1,-1){0.64}}
   \@ifstar{\@flSW}{\@@flSW}}
\def\@flSW #1{\put(-0.52,-0.48){\makebox(0,0)[br]{$#1$}}\end{picture}}
\def\@@flSW#1{\put(-0.48,-0.52){\makebox(0,0)[tl]{$#1$}}\end{picture}}
\def\flS{\begin{picture}(0,0)
   \put(    0,-0.2 ){\vector( 0,-1){0.60}}
   \@ifstar{\@flS}{\@@flS}}
\def\@flS  #1{\put(-0.03,-0.5 ){\makebox(0,0)[ r]{$#1$}}\end{picture}}
\def\@@flS #1{\put( 0.03,-0.5 ){\makebox(0,0)[ l]{$#1$}}\end{picture}}
\def\flSE{\begin{picture}(0,0)
   \put( 0.18,-0.18){\vector( 1,-1){0.64}}
   \@ifstar{\@flSE}{\@@flSE}}
\def\@flSE #1{\put( 0.48,-0.52){\makebox(0,0)[tr]{$#1$}}\end{picture}}
\def\@@flSE#1{\put( 0.52,-0.48){\makebox(0,0)[bl]{$#1$}}\end{picture}}
\def\capsa(#1,#2)#3{\put(#1,#2){\makebox(0,0){$#3$}}}
\def\diagr{\@ifnextchar [{\@diagr}{\@diagr[15ex]}}
\def\@diagr[#1](#2,#3){\begingroup
   \setlength{\unitlength}{#1}
   \begin{picture}(#2,#3)}
\def\enddiagr{\end{picture}
   \endgroup}

\makeatletter

\def\qed{\ifvmode\removelastskip\fi
{\unskip\nobreak\hfil\penalty50\hbox{}\nobreak\hfil
\hbox{\vrule height1.2ex width1.2ex}\parfillskip=0pt
\finalhyphendemerits=0 \par\smallskip}}

\def\dif{{\rm d}}
\def\deriv{\@ifnextchar[{\@deriv}{\@deriv[]}}
   \def\@deriv[#1]#2#3{\mathchoice%
{{\dif^{#1}#2\over\dif{#3}^{#1}}}{{\dif^{#1}#2/\dif{#3}^{#1}}}%
{{\dif^{#1}#2\over\dif{#3}^{#1}}}{{\dif^{#1}#2/\dif{#3}^{#1}}}}
\def\derpar#1#2{\mathchoice%
{{\partial#1\over\partial#2}}{{\partial#1/\partial#2}}%
{{\partial#1\over\partial#2}}{{\partial#1/\partial#2}}}

\def\restric#1#2{{\left. #1 \right|_{#2}}}

\def\secteqno{\@addtoreset{equation}{section}%
\def\theequation{\thesection.\arabic{equation}}}
\newcounter{subequation}
\def\thesubequation{\alph{subequation}}
\def\sneqnarray{\stepcounter{equation}\let\@currentlabel=\theequation
\setcounter{subequation}{1}
\def\@eqnnum{{\rm (\theequation.\thesubequation)}}
\global\@eqcnt\z@\tabskip\@centering\let\\=\@eqncr\let\@@eqncr=\@@sneqncr
$$\halign to \displaywidth\bgroup\@eqnsel\hskip\@centering
 $\displaystyle\tabskip\z@{##}$&\global\@eqcnt\@ne
 \hskip 2\arraycolsep \hfil${##}$\hfil
 &\global\@eqcnt\tw@ \hskip 2\arraycolsep $\displaystyle\tabskip\z@{##}$\hfil
  \tabskip\@centering&\llap{##}\tabskip\z@\cr}
\def\endsneqnarray{\@@sneqncr\egroup $$\global\@ignoretrue}
\def\@@sneqncr{\let\@tempa\relax
   \ifcase\@eqcnt \def\@tempa{& & &}\or \def\@tempa{& &}
   \else \def\@tempa{&}\fi
     \@tempa \if@eqnsw\@eqnnum\stepcounter{subequation}\fi
     \global\@eqnswtrue\global\@eqcnt\z@\cr}
\def\nobiblabels{\def\@lbibitem[##1]##2{\@bibitem{##2}}}
\makeatother
\def\ben{\begin{enumerate}}
\def\een{\end{enumerate}}
\def\beq{\begin{equation}}
\def\eeq{\end{equation}}
\def\bea{\begin{eqnarray}}
\def\eea{\end{eqnarray}}
\def\beann{\begin{eqnarray*}}
\def\eeann{\end{eqnarray*}}
\def\beasn{\begin{sneqnarray}}
\def\eeasn{\end{sneqnarray}}

\newtheorem{teor}{Theorem}
\newtheorem{prop}{Proposition}
\newtheorem{lem}{Lemma}
\newtheorem{cor}{Corollary}

\let\ds=\displaystyle
\def\buildord#1\over#2{\mathord{\mathop{\kern0pt #2}\limits^{#1}}}
\def\ir#1{{\buildord{\scriptscriptstyle\circ}\over #1}}

\def\mapping#1{\mathrel{\mathop{\longrightarrow}\limits^{#1}}}
\def\inmap#1{\mathrel{\mathop{\hookrightarrow}\limits^{#1}}}
\def\surmap#1{\mathrel{\mathop{\twoheadrightarrow}\limits^{#1}}}

\def\Real{{\bf R}}
\let\isom=\cong

\def\Ker{\mathop{\rm Ker}\nolimits}

\def\transp#1{{}^{t}\kern-.15em\relax#1}
\def\Tens^#1{\buildord #1\over\otimes }

\def\Img{\mathop{\rm Im}\nolimits}

\def\Cinfty{{\rm C}^\infty}

\def\feble#1{\mathrel{\mathop{\simeq}\limits_{#1}}}

\def\Tan{\mathrm{T}}
\def\Ver{{\rm V}}
\def\grad{\mathop{\rm grad}\nolimits}

\def\Sb{{\rm Sb}}

\def\fin{{\rm f}}
\def\rank{\mathop{\rm rank}\nolimits}

\def\calP{{\cal P}}
\def\calQ{{\cal Q}}
\let\eps=\varepsilon
\let\Ups=\Upsilon

\def\tanvec#1{{\partial \over \partial #1}}

\def\Vect{\mathfrak{X}}

\def\FD{\mathcal{F}}
\def\Fl{\mathrm{F}}

\def\proof{\noindent{\itshape Proof}.\quad}

\secteqno

\def\tabaddress#1{{\small\sffamily\slshape\begin{tabular}[t]{c}#1
\\[1.2ex]\end{tabular}}}
\def\UPCMAT{Departament de Matem\`atica Aplicada IV\\
   Universitat Polit\`ecnica de Catalunya\\
   Campus Nord UPC, edifici C3\\
   C.~Jordi Girona~1,
   08034 Barcelona\\
   Catalonia, Spain}

\title{\sffamily Regularity and symmetries of nonholonomic systems}

\author{\sffamily
Xavier Gr\`acia and Rub\'en Mart\'{\i}n
\\[1mm]
\tabaddress{\UPCMAT}
\\[2mm]
\small\sffamily\slshape emails: xgracia@mat.upc.es, rubenmg@mat.upc.es
}

\date{\sffamily 29 May 2004}

\begin{document}

\maketitle
\thispagestyle{empty}

\begin{abstract}
\parindent 0pt \noindent
Lagrangian systems with nonholonomic constraints
may be considered as
singular differential equations defined by some constraints
and some multipliers.
The geometry, solutions, symmetries and constants of motion
of such equations
are studied within the framework of
linearly singular differential equations.
Some examples are given; in particular
the well-known singular lagrangian of the
relativistic particle, which with the nonholonomic
constraint $v^2=c^2$ yields a regular system.

\bigskip
\itshape


MSC\,2000: 34A08, 70H
\quad

\end{abstract}

\section{Introduction}

The main goal of this paper is to study
the nonholonomic mechanical systems
within the framework of linearly singular
differential equations.

Nonholonomic mechanical systems, i.e.,
mechanical systems with non-integrable kinematic
constraints, have been discussed since the
last years of nineteenth century.
However, the geometric foundations for the
theory were given in \cite{VF-diff}.
Since then, several approaches have been
taken to deal with the subject, for instance,
a hamiltonian approach in \cite{BS-nonh},
a lagrangian  approach in \cite{LM-lag},
a more general Poisson framework in \cite{Mar-approaches},
or an approach based on a gauge independent formulation of
lagrangian and hamiltonian mechanics in \cite{MVB-nonh}.
Symmetries of these systems, as well as
reduction schemes derived from them, have also
been considered in the literature, see
\cite{Koi-red,BKMM-sym,KM-Poisson,CL-redrec,Mar-sym}.

A lagrangian system with nonholonomic constraints
may be considered, more generally, as
a singular differential equation defined by
some constraints and some multipliers:
$$
\dot x = g(x) + \sum_\mu u^\mu \, h_\mu(x) ,
\qquad
\phi_\alpha(x) = 0 .
$$
Such an equation can be described geometrically as a
{\itshape linearly singular differential equation},
that is, a differential equation
where the velocities are not isolated because of a linear
factor multiplying them:
$$
A(x) \dot x = b(x) .
$$
This is a special type of
{\itshape implicit differential equation}.
The idea of modelling mechanical systems as implicit
differential equations
is found in earlier papers by Tulczyjew
\cite{MT-ham,MMT-integrability},
and it has also been used to deal with nonholonomic constraints
\cite{Tul-constraints,ILMM-implicit}.

Linearly singular differential equations
were geometrically presented in
\cite{GP-unif, GP-gener}.
This general framework includes for instance the presymplectic
systems and the lagrangian formalism
\cite{GP-gener},
the higher order singular lagrangians
and their ``higher order differential equation'' conditions
\cite{GPR-higher},
as well as many other systems that appear in technological applications.
To solve the corresponding equation of motion
a consistency algorithm can be performed.
This algorithm is indeed a generalization of
the presymplectic constraint algorithm
\cite{GNH-pres}.

We will see that a system with constraints and multipliers,
and in particular any nonholonomic mechanical system,
can be described as a linearly singular system.
This implies that all the methods and results about these systems 
can be applied directly to nonholonomic systems. 

More precisely, the combination of two operations
that can be performed on linearly singular systems
---restriction to a subsystem and projection to a quotient---
can be applied to obtain what we call a generalized nonholonomic
system.
In particular,
we discuss the regularity, consistency and
equations of motion of these derived systems.

The symmetries of a linearly singular differential equation have
been studied in \cite{GP-sym}. In this paper we consider the
relation between the symmetries of a system with nonholonomic
constraints and the symmetries of its original unconstrained
system, both modelled on linearly singular differential equations.
We also study their constants of motion.

The paper is organized as follows.
In section 2 we give some
definitions and results regarding linearly singular differential
equations, their solutions and their symmetries.
In section 3 we introduce generalized nonholonomic systems
and discuss some of their properties.
Symmetries and constants of motion of
generalized nonholonomic systems are discussed in section~4.
In section 5 we show
how a lagrangian system with nonholonomic constraints can be
described in terms of a generalized nonholonomic systems.
The case of a relativistic particle is studied in section~6,
where we see that a nonholonomic constraint can convert a
singular lagrangian into a regular system.
Two additional examples are studied in section~7.
Finally, an appendix contains some auxiliary results formulated
within the framework of linear algebra.

\section{Previous results: linearly singular systems}
\label{sec-gener}

In this section we recall some definitions and results from
\cite{GP-unif, GP-gener, GP-sym}.

Let $M$ be a manifold.
An {\itshape implicit differential equation} on $M$ is defined by
a submanifold $D \subset \Tan M$.
A path $\xi \colon I \to M$ is a solution of this equation when
\beq
\dot\xi(I) \subset D.
\eeq
In coordinates, if the submanifold $D$ is described by some equations
$F=0$ and
the path $\xi$ is represented by some functions $x(t)$,
then the local expression of the implicit
differential equation is $F(x,\dot x)=0$.

We have a particular case when $D=X(M)$, with $X$ a vector field on~$M$.
Then $X$ defines an {\itshape explicit differential equation},
and $\xi$ is a solution iff
\beq
\dot\xi = X \circ \xi.
\eeq
Now the local expression is $\dot x=f(x)$.

A {\itshape linearly singular differential equation}
on $M$ is defined by a vector bundle $\pi\colon F \to M$,
a vector bundle morphism $A \colon \Tan M \to F$,
and a section $f\colon M \to F$ of $\pi$. A path
$\xi \colon I \to M$ is a solution when
\beq
\label{eqmov-xi}
A \circ \dot\xi = f \circ \xi,
\eeq
whose local expression is $A(x)\dot x=f(x)$,
with $A(x)$ a (singular, in general) matrix.

We denote by $(A\colon \Tan M \to F, f)$ the linearly singular system.
The following diagram shows all these data:
$$
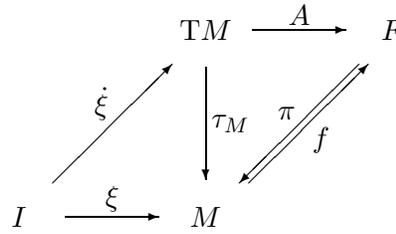
\diagr(2,1)
\capsa(0,0){I}
\capsa(1,0){M}
\capsa(1,1){\Tan M}
\capsa(0,0){\flE{\xi}}
\capsa(0,0){\flNE{\dot \xi}}
\capsa(1,1){\flS{\tau_M}}
\capsa(2,1){F}
\capsa(2,1){\flSW*{\pi}}
\capsa(1.05,0){\flNE*{f}}
\capsa(1,1){\flE{A}}
\enddiagr
$$
The associated implicit differential equation
is
\beq
D=A^{-1}(f(M)) \subset \Tan M.
\eeq
We say that the linearly singular differential
equation is {\itshape regular} when $A$ is
a vector bundle isomorphism. In this case,
the associated explicit differential equation
is given by the vector field $X=A^{-1}\circ f$.

The solutions of the system can be equivalently
described as integral curves of vector fields.
Let us remark that in general the solutions
are restricted to a submanifold $S \subset M$
because the equation (\ref{eqmov-xi}) may not have
solutions passing through every point $x \in M$.
Therefore, the equation of motion can be written as
an equation for a vector field $X$ and a
submanifold $S$:
\beq
\label{eqmov-X}
\left\{
  \begin{array}{l}
    X \mbox{ tangent to } S \\
    A \circ X \feble{S} f,
  \end{array}
\right.
\eeq
where the notation $\feble{S}$ means
equality at the points of~$S$.

A recursive algorithm can be applied to find the solutions
of a linearly singular differential equation.
Its first step is to note that, in order that a solution
passes through a point $x \in M$, it is necessary that
\beq
f(x) \in \Img A_x,
\eeq
so the solutions are necessarily contained in
the primary constraint subset
\beq
M_1=\{x\in M \mid f(x)\in \Img A_x\} ,
\eeq
which will be assumed to be a closed submanifold.
The tangency to~$M_1$ forces the initial system to be restricted to
$(A_1 \colon \Tan M_1 \to F_1, f_1)$,
where $A_1=\restric{A}{\Tan M_1}$,
$F_1=\restric{F}{M_1}$ and
$f_1=\restric{f}{M_1}$.
The algorithm follows recursively,
%
%
and, under some regularity assumptions at each step,
it ends with a final constraint submanifold $S$
such that $f(S) \subset \Img A_S$;
thus the system is consistent,
and the equation
\beq
A_S \circ X = f_S
\eeq
for a vector field $X$ tangent to $S$ has solutions.
Given a particular solution $X_\circ$, the set
of solutions of (\ref{eqmov-X}) is $X_\circ+\Ker A_S$.

\medskip

We finish this section by giving some definitions
and results about symmetries. 
A {\itshape symmetry of a linearly singular system}
$(A \colon \Tan M \to F,f)$ is a vector bundle
automorphism $(\varphi,\Phi)$ of
$\pi \colon F \to M$ such that
\beq
f= \Phi_{*}[f]:= \Phi \circ f \circ \varphi^{-1} ,
\qquad
A= \Phi_{*}[A]:= \Phi \circ A \circ {(\Tan \varphi)}^{-1} .
\label{symm}
\eeq
An {\itshape infinitesimal symmetry} of a linearly
singular system $(A \colon \Tan M \to F,f)$ is 
an infinitesimal automorphism $(V,W)$ of the vector bundle
$\pi \colon F \to M$ such that its flow
$(\Fl_{V}^{\varepsilon},\Fl_{W}^{\varepsilon})$ is
constituted by local symmetries of the
linearly singular differential equation.
The last property is equivalent to the following conditions:
\beq
\Tan f \circ V = W \circ f ,
\qquad
\Tan A \circ V^{T} = W \circ A ,
\label{infsymm}
\eeq
which are the infinitesimal version of~(\ref{symm}).

\section{Generalized nonholonomic systems}

\subsection*{The geometric setting}

Among the various operations that can be performed with
a linearly singular system
$(B \colon \Tan N \to G,g)$,
we are especially interested in
the subsystem defined on a submanifold
$j \colon M \hookrightarrow N$,
and the projection
$p \colon G \to G/G'$
to a quotient with respect to a vector subbundle $G' \subset G$:
$$
\diagr[12ex](5,1.2)
\capsa(0,0){N}
\capsa(0,1){\Tan N}
\capsa(0,1){\flS{}}
\capsa(1,1){G}
\capsa(1,1){\flSW*{}}
\capsa(0.07,0){\flNE*{g}}
\capsa(0,1){\flE{B}}

\capsa(2,0){M}
\capsa(2,1){\Tan M}
\capsa(2,1){\flS{}}
\capsa(3,1){\restric{G}{M}}
\capsa(3,1){\flSW*{}}
\capsa(2.07,0){\flNE*{g|_{M}}}
\capsa(2,1){\flE{\restric{B}{\Tan M}}}

\capsa(4,0){N}
\capsa(4,1){\Tan N}
\capsa(4,1){\flS{}}
\capsa(5.03,1){G/G'}
\capsa(5,1){\flSW*{}}
\capsa(4.07,0){\flNE*{p \circ g}}
\capsa(4,1){\flE{p \circ B}}
\enddiagr
$$

Suppose that the original system
admits solutions $Y$ on a submanifold $N_\fin \subset N$.
Then the subsystem on~$M$
has solutions on the submanifolds of $M \cap N_\fin$
over which a solution $Y$ of the initial system is tangent.
On the other hand,
the quotient system
has, in general, more solutions than the initial system:
if $Z$ is any vector field on $N$ tangent to $N_\fin$ with values
in $B^{-1}(G')$
then $Y+Z$ is a solution of the quotient system on~$N_\fin$;
there may also exist solutions defined on a submanifold
greater than~$N_\fin$.

It is well known that
the dynamics of systems with nonholonomic constraints
is a mixture of both constructions:
the presence of some constraints,
combined with a certain degree of arbitrariness
expressed through some multipliers.
This combination may result advantageous:
though in general $Y$ is not tangent to the submanifold~$M$,
it may happen that for some vector fields~$Z$
in $B^{-1}(G')$
one has solutions $Y+Z$
tangent to~$M$,
or at least to a ``big'' submanifold of~$M$.

In this paper we will call a
\textit{generalized nonholonomic system}
the linearly singular system
$(A \colon \Tan M \to F, f)$
defined from
$(B \colon \Tan N \to G, g)$
by
a \textit{constraint submanifold}
$M \subset N$ and
a \textit{subbundle
of constraint forces}
$G' \subset \restric{G}{M}$
as follows:
\begin{itemize}
\itemsep 0pt plus 1pt
\item
$F = (\restric{G}{M})/G'$,
\item
$A = p \circ \restric{B}{M} \circ \ir\Tan j$, and
\item
$f = p \circ \restric{g}{M}$,
\end{itemize}
where $p\colon \restric{G}{M} \to (\restric{G}{M})/G'$
is the projection to the quotient, 
and $\ir\Tan j$ denotes the tangent map of~$j$ with the image 
restricted to~$M$. 
All this is shown in the following diagram:
$$
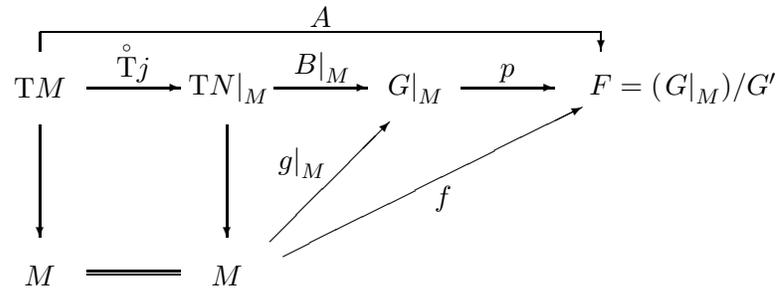
\diagr(3.25,1.5)
\capsa(0,0){M}
\capsa(0,1){\Tan M}
\capsa(0,1){\flS{}}
\capsa(0,1){\flE{\ir\Tan j}}
\put(0.25,0){\line(1,0){0.5}}
\put(0.25,0.02){\line(1,0){0.5}}
\capsa(1,0){M}
\capsa(1,1){\restric{\Tan N}{M}}
\capsa(1,1){\flS{}}
\capsa(2,1){\restric{G}{M}}
\capsa(1.05,0){\flNE{\restric{g}{M}}}
\capsa(1,1){\flE{\restric{B}{M}}}
\capsa(3,1){F \rlap{${} = (\restric{G}{M})/G'$}}
\capsa(2,1){\flE{p}}

\put(1.3,0.1){\vector(2,1){1.6}}
\put(2.11,0.48){\makebox(0,0)[tl]{$f$}}

\put(0,1.2){\line(0,1){0.1}}
\put(0,1.3){\line(1,0){3}}
\put(3,1.3){\vector(0,-1){0.1}}
\put(1.5,1.33){\makebox(0,0)[b]{$A$}}
\enddiagr
$$

\subsection*{Regularity and consistency}

Before discussing the equations of motion,
we want to study some general properties of the
generalized nonholonomic system
$(A \colon \Tan M \to F, f)$,
namely,
whether $A$ is surjective (we will also say that the system is surjective)
or bijective (the system is regular),
or the equation $A \circ X = f$ is everywhere consistent.

Let us denote
$$
H = B^{-1}(G') \subset \restric{\Tan N}{M} ,
$$
which is a vector subbundle whenever the morphism $B$ has constant rank.

\begin{prop}
With the preceding notations,
the generalized nonholonomic system is surjective iff
$$
B(\Tan M) + G' = \restric{G}{M} .
$$
Assuming that the original system is surjective,
the nonholonomic system is surjective iff
$$
\Tan M + H = \restric{\Tan N}{M} ,
$$
and it is regular iff in addition
$$
\Tan M  \cap  H = \{0\} .
$$
\label{reg-cons}
\end{prop}
\proof
We want to decide whether
$A = p \circ \restric{B}{M} \circ \ir\Tan j$
(the composition of an inclusion, a morphism and a projection)
is surjective or injective,
and this is given by lemma~1 in the appendix.
\qed

The preceding result could be refined also in the case where
$B$ is injective,
but this does not seem so interesting.
As an immediate consequence, we have:
\begin{cor}
Suppose that the original system is surjective
(or, more particularly, regular).
Then the generalized nonholonomic system is regular iff
$$
\restric{\Tan N}{M} = \Tan M  \oplus  H .
$$
\qed
\label{cor-dirsum}
\end{cor}

These relations can be given a more concrete form
in terms of constraints and frames.
Consider a local basis
$(\Gamma_\mu)_{1\leq \mu \leq m_\circ}$
of sections for the subbundle~$H \subset \restric{\Tan N}{M}$
(they are vector fields in~$N$, but defined only on~$M$).
Consider also
a set of $a_\circ$ constraints $\phi^\alpha$,
linearly independent at each point,
that locally define the submanifold
$M \subset N$.
Finally, consider the matrix
\beq
D^{\alpha}_{\,\mu} =
\langle \restric{\dif\phi^\alpha}{M} , \Gamma_\mu \rangle =
\Gamma_\mu \cdot \phi^\alpha ,
\label{matrix}
\eeq
whose elements are functions on~$M$.

\begin{prop}
With the preceding notations,
\ben
\item
$\Tan M \cap H = 0$
iff
$\rank(D^{\alpha}_{\,\mu}) = m_\circ$.
\item
$\Tan M + H = \restric{\Tan N}{M}$
iff
$\rank(D^{\alpha}_{\,\mu}) = a_\circ$.
\item
$\Tan M \oplus H = \restric{\Tan N}{M}$
iff
$(D^{\alpha}_{\,\mu})$ is a square invertible matrix.
\een
\end{prop}
\proof
It is a consequence of lemma~3 in the appendix,
since the $\restric{\dif\phi^\alpha}{M}$ constitute a basis for the
annihilator of $\Tan M$ in $(\restric{\Tan N}{M})^*$.
\qed

The connection of such a matrix with the notion of
regularity and consistency of a constrained system was already noted
in 
\cite{CR-multipliers, LM-lag}.

\subsection*{Equations of motion}

From the definition of the generalized nonholonomic system
$(A \colon \Tan M \to F, f)$,
it is clear that a path $\xi \colon I \to N$ is
a solution of the equation of motion iff it is contained in~$M$
and
\beq B \circ \dot\xi - g \circ \xi \in G' .
\label{nh-path}
\eeq
If some sections $\Delta_\nu$ constitute a frame for~$G'$,
then this equation can be written as
\beq
B \circ \dot\xi = g \circ \xi + \sum_\nu v^\nu \,\Delta_\nu \circ \xi ,
\label{nh-path'}
\eeq
for some multipliers $v^\nu(t)$.

In the same way,
for a submanifold $S \subset M$ and a vector field $X$ on~$M$
tangent to~$S$,
the equation of motion
$A \circ X \feble{S} f$
can be written as
\beq
B \circ X - g \mathrel{\mathop{\in}_{S}} G' ,
\label{nh-vf}
\eeq
where the equation must only hold on the points of~$S$.
This equation may be also written as
\beq
B \circ X  \feble{S}  g + \sum_\nu v^\nu \Delta_\nu ,
\label{nh-vf'}
\eeq
for some multipliers $v^\nu(x)$.

Of course,
we can apply the constraint algorithm
to find the solutions of this linearly singular system.
However,
there is an alternative way to solve the problem
when the original problem is regular, or at least consistent.
Under this hypothesis,
let $Y$ be a vector field on~$N$,
solution of the equation of motion of the linearly singular system
$(B \colon \Tan N \to G,g)$:
$$
B \circ Y = g .
$$
(For most applications the original system is regular,
and then the unique solution of this equation is the vector field
$
Y = B^{-1} \circ g
$.)

Using~$Y$,
the equations of motion become
\beq
\dot\xi - Y \circ \xi \in H
\label{nh-path''}
\eeq
for a path $\xi$ in~$M$, and
\beq
X - Y \mathrel{\mathop{\subset}_{S}} H ,
\label{nh-vf''}
\eeq
for a vector field $X$ on $M$ that should be tangent to~$S$.

These equations can be expressed in a more concrete form
in terms of the local basis
$(\Gamma_\mu)$
of sections for the subbundle~$H \subset \restric{\Tan N}{M}$:
\beq
\dot\xi = Y \circ \xi + \sum_\mu u^\mu \,\Gamma_\mu \circ \xi ,
\label{nh-path'''}
\eeq
for some functions~$u^\mu(t)$, and
\beq
X \feble{S} Y + \sum_\mu u^\mu \,\Gamma_\mu ,
\label{nh-vf'''}
\eeq
for some functions~$u^\mu$ on~$M$.

Let us examine whether this last equation has solutions.
The requirement for $X$ of being tangent to $M$ is
$X \cdot \phi^\alpha \feble{M} 0$,
which reads
\beq
\sum_\mu D^{\alpha}_{\,\mu} \, u^\mu + Y \cdot \phi^\alpha \feble{M} 0 ,
\label{compat'}
\eeq
where $(D^{\alpha}_{\,\mu})$ is the matrix defined by (\ref{matrix}).
From this it is clear that
the generalized nonholonomic system is regular
iff
the matrix $(D^{\alpha}_{\,\mu})$ is invertible on~$M$,
and in this case the equation (\ref{compat'})
directly determines the functions $u^\mu$
that give the solution $X$ expressed in (\ref{nh-vf'''}).
More generally,
the nonholonomic system has solutions
if the matrix $(D^{\alpha}_{\,\mu})$ has rank~$a_\circ$.

Geometrically, the decomposition
$\restric{\Tan N}{M} = \Tan M \oplus H$
stated in Corollary~\ref{cor-dirsum}
has two associated projectors $\calP$, $\calQ$.
Writing
$Y = \calP \circ Y + \calQ \circ Y$ on~$M$,
the following result is clear:
\begin{prop}
\label{prop-proj} 
With the preceding notations,
if
the original system is consistent, with a solution~$Y$,
and the generalized nonholonomic system is regular, with solution~$X$,
the latter can be obtained as
\beq
X = \calP \circ \restric{Y}{M} .
\eeq
\qed
\end{prop}

Such projectors were studied, in the context of
nonholonomic lagrangian systems, in 
\cite{LM-lag}.

\section{Symmetries and constants of motion}


Let us consider a generalized nonholonomic
system $(A \colon \Tan M \to F,f)$, obtained
from a linearly singular system
$(B \colon \Tan N \to G,g)$ by means of a
restriction to a submanifold $M \subset N$
and a projection to the quotient
$p \colon \restric{G}{M} \to (\restric{G}{M})/G'$,
where $G' \subset \restric{G}{M}$
is a vector subbundle.

Recall the definitions of symmetry and infinitesimal symmetry 
given in section~2.
Our aim is to study the relation between
the symmetries of the original linearly singular system on $N$ 
and the symmetries 
of the generalized nonholonomic system on~$M$.
In the next proposition, we give sufficient conditions on 
a symmetry of the original system in order to define 
a symmetry of the constrained system:
\begin{prop}
\label{prop-sym}
Let $(\psi,\Psi)$ be a symmetry of
$(B \colon \Tan N \to G,g)$.
Suppose that
$\psi$ leaves the submanifold $M \subset N$
invariant,
and $\Psi$ leaves the subbundle $G' \subset \restric{G}{M}$ invariant.
Then
$(\varphi,\Phi)$,
where $\varphi=\restric{\psi}{M}$,
and $\Phi \colon (\restric{G}{M})/G' \to (\restric{G}{M})/G'$
is the map induced on the quotient from~$\Psi$,
is a symmetry of $(A \colon \Tan M \to F,f)$.
\end{prop}
\proof
We have
$$
A \circ \Tan \varphi=
p \circ B \circ \Tan j \circ \Tan (\restric{\psi}{M})=
p \circ B \circ \Tan \psi \circ \Tan j=
p \circ \Psi \circ B \circ \Tan j=
\Phi \circ p \circ B \circ \Tan j=
\Phi \circ A,
$$
and
$$
f \circ \varphi=
p \circ g \circ \restric{\psi}{M}=
p \circ \Psi \circ \restric{g}{M}=
\Phi \circ p \circ \restric{g}{M}=
\Phi \circ f,
$$
so the two conditions for being a symmetry are satisfied.
\qed

We can obtain a similar result for infinitesimal symmetries,
by making use of their infinitesimal characterization
(\ref{infsymm}):
\begin{prop}
Let $(V,\bar{V})$ be an infinitesimal symmetry
of $(B \colon \Tan N \to G,g)$.
Suppose that $V$ is tangent to the submanifold $M\subset N$,
and $\bar V$ is tangent to the subbundle $G'\subset \restric{G}{M}$.
Then $(U,\bar U)$,
where $U=\restric{V}{M}$ and
$\bar{U} \colon (\restric{G}{M})/G' \to \Tan ((\restric{G}{M})/G')$
is the vector field induced on the quotient from~$\bar{V}$,
is an infinitesimal symmetry of $(A \colon \Tan M \to F,f)$.
\end{prop}
\proof
The proof runs as in proposition \ref{prop-sym}:
$$
\Tan f \circ U =
\Tan p \circ \Tan g \circ \restric{V}{M} =
\Tan p \circ \bar{V} \circ \restric{g}{M}=
\bar{U} \circ p \circ \restric{g}{M}=
\bar{U} \circ f,
$$
\beann
\Tan A \circ U^T & = &
\Tan p \circ \Tan B \circ \Tan (\Tan j) \circ \restric{(V^{T})}{\Tan M}=
\Tan p \circ \Tan B \circ V^{T} \circ \Tan j = \\
 & = & \Tan p \circ \bar{V} \circ B \circ \Tan j =
\bar{U} \circ p \circ B \circ \Tan j=
\bar{U} \circ A.
\eeann
\qed

\medskip

We now consider constants of motion.
Suppose that the original system has a solution $Y \in \Vect(N)$, 
and let us consider a function $h \in \Cinfty (N)$ 
such that $Y \cdot h=0$. 
Under which conditions is $\restric{h}{M}$
a constant of motion of the generalized nonholonomic system?

Suppose that both the original system and the nonholonomic system
are regular, so that
$\restric{\Tan N}{M} = \Tan M \oplus H$; 
let $\calP$ be the projector to the first factor, 
which, according to Proposition~\ref{prop-proj}, 
relates the dynamics of both systems as 
$X = {\cal P} \circ Y$. 
Then we have a simple characterization:
\begin{prop}
\label{prop-constants}
With the preceding hypothesis,
write $X = Y-\Gamma$,
where $\Gamma$ is a section of~$H \subset \restric{\Tan N}{M}$.
Let $h$ be a constant of motion of the unconstrained
system.
Then $\restric{h}{M}$ is a constant of motion of the generalized
nonholonomic system iff $\Gamma \cdot h = 0$.
\end{prop}
\proof
It is straightforward:
$$
X \cdot h =
(Y-\Gamma) \cdot h =
Y \cdot h - \Gamma \cdot h .
$$
(Note that $Y$ and~$\Gamma$, considered as sections of
$\restric{\Tan N}{M}$, map functions on $N$ to functions
on~$M$.) \qed

\section{Lagrangian systems with nonholonomic constraints}

\def\Leg{\FD L}

In this section we will show that
the dynamics of
a lagrangian system with nonholonomic constraints
(the {\it nonholonomic mechanics}\/)
falls into the class of
generalized nonholonomic systems of section~3.

We begin by considering a configuration manifold~$Q$,
its tangent bundle $\Tan Q$,
and a lagrangian function
$L \colon \Tan Q \to \Real$.
The lagrangian mechanics may be described as the
linearly singular system
$(\widehat\omega_L \colon \Tan(\Tan Q) \to \Tan^*(\Tan Q), \dif E_L)$.
$$
\diagr[12ex](1,1.2)
\capsa(0,0){\Tan Q}
\capsa(-0.1,1){\Tan(\Tan Q)}
\capsa(0,1){\flS{}}
\capsa(1.1,1){\Tan^*(\Tan Q)}
\capsa(1,1){\flSW*{}}
\capsa(0.07,0){\flNE*{\dif E_L}}
\capsa(0,1){\flE{\widehat\omega_L}}
\enddiagr
$$
Here $E_L$ is the lagrangian energy and $\omega_L$ is the
Lagrange's 2-form. Though we do not want to dwell on these well
known objects, some properties of $\omega_L$ and the vertical
endomorphism will be needed later, so let us briefly recall them.
See \cite{Car-theory} for more details.

First, we have the vertical endomorphism $J$ of $\Tan(\Tan Q)$,
whose kernel and image are the vertical subbundle $\Ver(\Tan Q)$.
Its transposed morphism is an endomorphism
$\transp{J}$ of $\Tan^*(\Tan Q)$,
whose kernel and image are $\Sb(\Tan Q)$, 
the bundle of semibasic forms.
This is used to define the Lagrange's
1-form
$\theta_L = \transp{J} \circ \dif L$
and 2-form
$\omega_L = -\dif \theta_L$
on $\Tan Q$.

From now on we consider the case of the lagrangian being regular,
which amounts to $\omega_L$ being a symplectic form.
Then,
it induces a vector bundle isomorphism
$\widehat\omega_L \colon \Tan(\Tan Q) \to \Tan^*(\Tan Q)$
mapping vertical vectors to semibasic forms,
thus yielding an isomorphism
$\Ver(\Tan Q) \mapping{\isom} \Sb(\Tan Q)$.

It is well known that the lagrangian dynamics on $\Tan Q$ 
is described by the only vector field $X_L$ solution of
$$
\widehat\omega_L \circ X_L = \dif E_L .
$$
Note that it is a second-order vector field.

Now let us introduce the nonholonomic constraints, which define a
submanifold $M \inmap{j} \Tan Q$ of dimension~$m$. We will
consider only the case where this submanifold restricts the
velocities, not the configuration coordinates. In a more formal
way, this is described by the conditions given in the next
proposition:

\begin{prop}
\label{prop-subm}
Let $M \subset \Tan Q$ be a submanifold.
The following conditions are equivalent:
\ben
\item
The projection $M \to Q$
(restriction of the tangent bundle projection $\tau_Q \colon \Tan Q \to Q$)
is a submersion.
\item
$(\Tan M)^\vdash \cap \Sb(\Tan Q)|_M = 0$.
\item
The submanifold $M \subset \Tan Q$ can be locally described
by the vanishing of some constraints $\phi^i$
whose fibre derivatives $\FD \phi^i$
are linearly independent at each point of~$M$.
\item
The submanifold $M \subset \Tan Q$ can be locally described
by the vanishing of some constraints $\phi^i$
such that the 1-forms $\Delta^i = \transp{J} \circ \dif\phi^i$
are linearly independent at each point of~$M$.
\een
\end{prop}

In coordinates, these conditions mean that
$\ds \left( \derpar{\phi^i}{v^k} \right)$ has maximal rank.
\qed

Note that under the preceding conditions
the image $\tau_Q(M)\subset Q$ is an open submanifold,
and so we can replace $Q$ with this submanifold.
So, from now on,
{\itshape
we assume that the projection $M \to Q$ is a \emph{surjective} submersion.
}

Now we will consider the following vector bundles:
\beann
     &  & \Tan M \subset \restric{\Tan(\Tan Q)}{M} ,
\\
     &  & (\Tan M)^\vdash \subset \restric{\Tan^*(\Tan Q)}{M} ,
\\
G'   &:=& \transp{J}((\Tan M)^\vdash) \subset \restric{\Sb(\Tan Q)}{M} ,
\\
H    &:=& \widehat\omega^{-1}(G') \subset \restric{\Ver(\Tan Q)}{M} .
\eeann
Suppose that $M \subset \Tan Q$
is defined by the vanishing of some independent constraints~$\phi^i$ 
as in the preceding proposition. 
Then $(\Tan M)^\vdash$ is spanned by the $\restric{\dif \phi^i}{M}$.
We denote by $\Delta^i$ and $\Gamma_i$ their corresponding images
in $G'$ (through $\transp{J}$)
and $H$ (through $\widehat\omega_L^{-1}$).
The following diagram shows all these objects:
$$
\diagr(3,1.2)
\put(0.4,0){\makebox(0,0)[r]{$\langle \Gamma_i \rangle = H
\hookrightarrow{}$}}
\put(0.4,1){\makebox(0,0)[r]{$\Tan M \hookrightarrow {}$}}
\put(2.6,0){\makebox(0,0)[l]{${}\hookleftarrow G' = \langle \Delta^i
\rangle$}}
\put(2.6,1){\makebox(0,0)[l]{
${}\hookleftarrow (\Tan M)^\vdash = \langle \restric{\dif \phi^i}{M}
\rangle$}}
\capsa(0.8,0){\restric{\Ver(\Tan Q)}{M} }
\capsa(2.2,0){\restric{\Sb(\Tan Q)}{M} }
\capsa(0.8,1){\restric{\Tan(\Tan Q)}{M} }
\capsa(2.2,1){\restric{\Tan^*(\Tan Q)}{M} }
\capsa(1,0){\flE{\widehat\omega}}
\capsa(1,1){\flE{\widehat\omega}}
\capsa(0.9,1){\flS{J}}
\capsa(2.1,1){\flS{\transp{J}}}
\capsa(0.85,0){\flN{}}
\capsa(2.05,0){\flN{}}
\enddiagr
$$

So we have two subbundles $\Tan M, H \subset \restric{\Tan(\Tan Q)}{M}$.
We have 
$\rank \Tan M = m$ 
and
$\rank (\Tan M)^\vdash = n-m$; 
the conditions in Proposition~\ref{prop-subm} imply also that
$\rank H = \rank G' = n-m$.

\begin{teor}
The nonholonomic mechanics defined by the lagrangian $L$ 
and the constraint submanifold $M \subset \Tan Q$ 
is the generalized nonholonomic system defined 
from the lagrangian mechanics 
$(\widehat\omega_L \colon \Tan(\Tan Q) \to \Tan^*(\Tan Q), \dif E_L)$ 
by the constraint submanifold $M \subset \Tan Q$  
and the subbundle of constraint forces 
$G' = \transp{J}((\Tan M)^\vdash) \subset \restric{\Tan^*(\Tan Q)}{M}$.
\end{teor}

$$
\diagr(4,1.2)
\capsa(0,0){M}
\capsa(0,1){\Tan M}
\capsa(0,1){\flS{}}
\capsa(0,1){\flE{\ir\Tan j}}
\put(0.25,0){\line(1,0){0.5}}
\put(0.25,0.02){\line(1,0){0.5}}
\capsa(1.05,0){M}
\capsa(1.1,1){\restric{\Tan(\Tan Q)}{M}}
\capsa(1.05,1){\flS{}}
\capsa(1.15,0){\flNE*{\restric{\dif E_L}{M}}}
\capsa(1.25,1){\flE{\restric{\widehat\omega}{M}}}
\capsa(2.4,1){\restric{\Tan^*(\Tan Q)}{M}}
\capsa(2.6,1){\flE{}}
\capsa(3.9,1){\restric{\Tan^*(\Tan Q)}{M}/G'}


\enddiagr
$$

\proof
The equation of motion for a path
$\xi = \dot\gamma$
such that $\xi(t) \in M$ is
\beq
\dot\xi = X_L \circ \xi + \sum_i u^i \,\Gamma_i \circ \xi .
\eeq
Instead, let us write the equations of motion for vector fields:
according to (\ref{nh-vf'}),
for a second-order vector field $X$ on $\Tan Q$,
tangent to~$M$, the equation is
\beq
i_X \omega_L  \feble{S}  \dif E_L + \sum_i u^i \Delta_i ,
\label{nh-lag}
\eeq
or, according to (\ref{nh-vf'''}),
\beq
X  \feble{S}  X_L + \sum_i u^i \Gamma_i .
\eeq
But in coordinates equation (\ref{nh-lag}) reads 
$$
\derpar{L}{q} - \deriv{}{t} \left( \derpar{L}{v} \right) =
\sum_i u^i \derpar{\phi_i}{v} ,
$$
which is the equation of motion of the nonholonomic mechanics 
defined from $L$ and the constraints
---see for instance 
\cite{Arn-dynsyst3}.
\qed

If, in addition to
$(\Tan M)^\vdash \cap \restric{\Sb(\Tan Q)}{M} = 0$,
we have $\Tan M \cap H = 0$,
then 
$\restric{\Tan(\Tan Q)}{M} = \Tan M \oplus H$,
and so there is a unique solution~$X$ of the equation of motion,
which can be obtained from $Y$ through the projector to $\Tan M$
as described by Proposition~\ref{prop-proj}.

\subsection*{The case of a singular lagrangian}

The preceding method can be conveniently adapted
if the lagrangian is singular.
Of course, one can not use the direct sum decomposition.
However,
the formulation of the nonholonomic dynamics
as a quotient system on a submanifold remains unchanged,
except that the second-order condition
is not automatically satisfied by $X$ and must be imposed
as an additional equation for it:
$$
J \circ X \feble{M} \Delta_{\Tan Q} .
$$

This condition may be included
in the equation of motion of the nonholonomic dynamics
in the same way as can be done with the lagrangian dynamics,
using the time-evolution operator $K$ of lagrangian dynamics
\cite{BGPR-equiv}
\cite{GP-K}.
With it, 
the lagrangian dynamics is the linearly singular system
$$
\diagr[12ex](1,1.2)
\capsa(0,0){\Tan Q}
\capsa(-0.1,1){\Tan(\Tan Q)}
\capsa(0,1){\flS{}}
\capsa(1.5,1){\Tan Q \!\times_{\Leg}\! \Tan(\Tan^*Q)}
\capsa(1,1){\flSW*{}}
\capsa(0.05,0){\flNE*{\ir K}}
\capsa(0,1){\flE{\ir\Tan(\Leg)}}
\enddiagr
$$
where $\Leg \colon \Tan Q \to \Tan^*Q$
is the Legendre's transformation (fibre derivative) of~$L$.
In terms of vector fields, 
the lagrangian dynamics is thus defined by the equation 
$$
\Tan(\Leg) \circ X \feble{} K .
$$
Then, it is readily seen that
the nonholonomic equation of motion can be written 
\beq
\Tan(\Leg) \circ X \feble{} K - \sum_i u^i \Ups^{\phi_i} .
\eeq
Here $\Ups^\phi$ is a certain vector field along~$\Leg$,
which is defined from the fibre derivative of a function
$\phi \colon \Tan^*Q \to \Real$
---see 
\cite{GP-struc} 
for details.

\section{Relativistic particle with a nonholonomic constraint}

In this section we study the motion of a relativistic particle
as a nonholonomic constrained system.
We will consider two possible lagrangian functions, 
a regular one 
(deeply studied in~\cite{KM-rel}) 
and a singular one. 

Let us consider a particle with mass~$m$ and charge~$e$ moving in spacetime.
We model spacetime as a 4-dimensional manifold~$Q$, endowed with a
metric tensor~$g$ of signature $(1,3)$.
Suppose furthermore that the particle is subject to
the action of an electromagnetic field~$F=\dif A$,
where $A \in \Omega^1(Q)$,
and a potential~$U \in \Cinfty (Q)$.

Recall that there are some relevant objects
associated with the metric~$g$, namely,
the isomorphism~$\widehat g\colon\Tan Q \to \Tan^*Q$
(we will denote $X^{\flat}=\widehat g\circ X)$),
the Levi-Civita connection~$\nabla$,
the differential forms~$\theta_g=\widehat g^*(\theta_Q)\in\Omega^1(\Tan Q)$
and $\omega_g=\widehat g^*(\omega_Q)=-\dif\theta_g\in\Omega^2(\Tan Q)$,
the energy~$E_g(u_q)=\frac{1}{2}g(u_q,u_q)\in\Cinfty (\Tan Q)$, and
the geodesic vector field~$S_g$, which satisfies
$i_{S_g} \omega_g = \dif E_g$.
We denote $v=\sqrt{2E_g}$.

We will study two different lagrangian functions, namely
$$
L_1(u_q) =
-mc \,g(u_q,u_q)^{1/2} - {e \over c} \langle A(q),u_q \rangle - U(q) ,
$$
and
$$
L_2(u_q) =
-{1 \over 2}m \,g(u_q,u_q) - {e \over c} \langle A(q),u_q \rangle - U(q) .
$$
Forgetting the potential, 
$L_1$ is the singular lagrangian commonly used in relativistic mechanics 
to describe a particle in an electromagnetic field; 
it is defined only on the open set of time-like vectors of $\Tan Q$. 
The lagrangian $L_2$ appears in~\cite{KM-rel}.
Our aim is to compare both systems,
and to introduce the nonholonomic constraint $v^2 = c^2$ to them. 

The lagrangians $L_1$ and $L_2$ have, respectively,
associated Lagrange's 1-forms
$\theta_1 = -{mc \over v}\theta_g -{e \over c}\tau^{*}_{Q} A$ and
$\theta_2 = -m\theta_g -{e \over c}\tau^{*}_{Q} A$;
the Lagrange's 2-forms are
$\omega_1 = -{mc \over v}\omega_g
            -{c \over v^2}\dif v \wedge \theta_g
            +{e \over c}\tau^{*}_{Q} F$ and
$\omega_2 = -m\omega_g + {e \over c}\tau^{*}_{Q} F$;
and the lagrangian energies are
$E_1 = U$ and $E_2 = -{1\over 2}mv^2 + U$.


The symplectic formulation
of the equations of motion
for the lagrangians
$L_1$ and $L_2$ are, respectively,

\beq
i_X\omega_1 = \dif E_1,
\label{eqmot1}
\eeq
and
\beq
i_X\omega_2 = \dif E_2,
\label{eqmot2}
\eeq
for second-order vector fields $X$.
For any 2-form~$\omega$, we will also denote $i_X\omega$ by
$\widehat\omega(X)$.

It is worth writing down the Euler--Lagrange equations of motion
for a path~$\gamma$, which are, for lagrangians
$L_1$ and $L_2$:
\beq
{mc \over g(\dot\gamma,\dot\gamma)^{1/2}}
\left( (\nabla_t \dot\gamma)^\flat -
{g(\dot\gamma,\nabla_t \dot\gamma) \over g(\dot\gamma,\dot\gamma)}
\dot\gamma^\flat\right ) +
{e \over c} i_{\dot\gamma}F - \dif U = 0,
\label{ELeq1}
\eeq
and
\beq
m(\nabla_t \dot\gamma)^\flat + {e \over c} i_{\dot\gamma}F - \dif U = 0.
\label{ELeq2}
\eeq

Let us now consider equations~(\ref{eqmot1}) and~(\ref{eqmot2}).

As $\widehat\omega_1$ is not surjective, equation~(\ref{eqmot1})
could have no solutions.
We denote by
$\Delta = \dot q^i\tanvec{\dot q^i}$
the Liouville vector field,
$T=\dot q^i\tanvec{q^i}$
the natural vector field along $\tau_Q$,
and
$\xi^{\vee}$ the vertical lift of a vector field
$\xi \colon \Tan Q \to \Tan Q$
along $\tau_Q$.
We have that
$\Ker \omega_1=\langle \Delta,
\Sigma\rangle$,
where
\beq
\Sigma = S_g-{ev\over mc^2}((i_T F)^{\sharp})^{\vee} .
\eeq

We can see that
$\widehat\omega_1 ({v\over mc}(\grad U)^{\vee}) = \dif U -
({1\over v^2}i_T\dif U)\theta_g$ and that
$\theta_g\not\in\Img\widehat\omega_1$. Therefore
equation (\ref{eqmot1}) has solutions if and only if
$i_T\dif U=0$, that is, the potential~$U$ is constant,
which, in practice, is the same as taking $U$ equal to $0$.

Since $\Sigma$ is a second-order vector field, in absence of
potential the solutions of equation~(\ref{eqmot1}) are $X_1 =
\Sigma + \mu\Delta$, where $\mu$ is an arbitrary function. If, in
addition, there is no electromagnetic field, then the solutions
are $S_g+\mu\Delta$, and their integral curves are reparametrized
geodesics.

On the other hand, equation~(\ref{eqmot2}) is regular,
and its solution is
\beq
X_2 = S_g + {1\over m}(\grad U)^{\vee}-{e\over mc}((i_T F)^{\sharp})^{\vee}.
\eeq
This can be proved making use of the relations
$i_{Z^{\vee}}\omega_g=-\tau^{*}_{Q}(Z^{\flat})$
for vector fields~$Z$ along $\tau_{Q}$, and
$i_S (\tau^*_Q F) = \tau^*_Q (i_T F)$.
In this case, in absence of electromagnetic field and potential,
the solutions are the geodesics of $g$.

Now we introduce the nonholonomic constraint
\beq
\phi(u_q) := g(u_q,u_q)-c^2 = 0 ,
\eeq
which defines a submanifold~$M \subset \Tan Q$.

The subbundle of constraint forces is
$\restric{\langle \transp{J}(\dif\phi)\rangle}{M} =
 \restric{\langle \theta_g\rangle}{M}$,
therefore, according to equation~(\ref{nh-lag}),
the equations of motion for both lagrangians become
\beq
i_X \omega_1 \feble{M} \dif E_1 + \lambda\theta_g,
\label{eqcon1}
\eeq
and
\beq
i_X \omega_2 \feble{M} \dif E_2 + \lambda\theta_g,
\label{eqcon2}
\eeq
for second-order vector fields $X$ tangent to~$M$.

Note that if a path~$\gamma$ satisfies the constraint
then it also satisfies the equation
$\ds
0 = \deriv{}{t} g(\dot\gamma,\dot\gamma) =
2 g(\dot\gamma,\nabla_t \dot\gamma)
$, 
so looking at equations~(\ref{ELeq1}) and~(\ref{ELeq2}) 
we realize that
the two {\it constrained}\/ systems have the {\it same}\/  
equations of motion:
\beq
\left\{ \begin{array}{l}
\ds
m (\nabla_t \dot\gamma)^\flat + {e \over c} i_{\dot\gamma}F - \dif U =
\lambda\dot\gamma^\flat ,
\\
g(\dot\gamma,\dot\gamma)= c^2 .
\end{array}
\right.
\eeq
The multiplier~$\lambda$ can be found by contracting the equation
with~$\dot\gamma$, which gives 
$\ds\lambda = -\frac{1}{c^2} i_{\dot\gamma} \dif U$.

We are going to see this equivalence of the solutions of both 
Euler--Lagrange equations by computing the solutions of 
equations~(\ref{eqcon1}) and~(\ref{eqcon2}).

First let us analyse equation~(\ref{eqcon2}).
From 
$\Delta \cdot \phi = 2v^2 \feble{M} 2c^2 \not= 0$ 
and 
$i_\Delta \omega_2 = m\theta_g$, 
it follows that 
$
\Tan M \oplus \widehat\omega_2^{-1}
(\restric{\langle\theta_g\rangle}{M}) = \restric{(\Tan Q)}{M}
$,
so, by proposition~\ref{reg-cons}, the system is regular. 
Its solution is $X=X_2+{\lambda\over m}\Delta$, 
where the multiplier~$\lambda$ is found by imposing that 
$X$ is tangent to~$M$: 
\beq 
0 = X \cdot \phi = 
X_2 \cdot \phi + {\lambda \over m} \Delta \cdot \phi 
\feble{M} 
{2\over m} i_T\dif U + 2{\lambda\over m} c^2. 
\eeq 
Therefore, the solution of the second system is \beq
X=S_g + {1\over m} (\grad U)^\vee - \frac{e}{mc} ((i_T
F)^{\sharp})^{\vee} -{1\over mc^2}(i_T \dif U)\Delta. 
\eeq

Now let us analyse equation (\ref{eqcon1}). 
Since 
$Y={1\over m} (\grad U)^\vee -{1\over mc^2}(i_T \dif U)\Delta$ 
is a vector field tangent to~$M$ and 
$\widehat\omega_1 (Y) \feble{M} 
\dif U - ({1\over c^2}i_T\dif U)\theta_g$, 
the system is consistent.
We can see that
\beq
\Tan M \cap
\widehat\omega_1^{-1}(\restric{\langle\theta_g\rangle}{M})=
\Tan M \cap \Ker\widehat\omega_1 =
\restric{\langle\Sigma\rangle}{M},
\eeq
so the system is not regular.
Then, the solutions of the equation are $Y+\mu\Sigma$.
Since $Y$ is vertical, in order 
to be a second-order vector field
the function $\mu$ must be equal to one, so
the solution is $Y+\Sigma\feble{M} X$, 
exactly the same as for the lagrangian $L_2$.

\section{Examples}

\subsection*{Example 1}

Consider the differential equation on $N=\Real^2$
defined by the vector field $Y=\tanvec{x}+y \tanvec{y}$.
We restrict this system to a generalized nonholonomic
one by means of the construction of section~3,
taking the submanifold $M=\Real \times \{a\}\subset N$ and the
subbundle $C=\langle x \tanvec{x} + \tanvec{y}\rangle
\subset \restric{\Tan N}{M}$.

In this case $\restric{\Tan N}{M}=\Tan M \oplus C$
and the projectors associated with this decomposition are
$$
\begin{array}{llll}
{\cal P} \colon & \tanvec{x} &
\longmapsto & \tanvec{x} 
\\
 & \tanvec{y} &
\longmapsto & -x \tanvec{x} ,
\end{array}
$$
$$
\begin{array}{llll}
{\cal Q} \colon & \tanvec{x} &
\longmapsto & 0 
\\
& \tanvec{y} &
\longmapsto & x \tanvec{x} + \tanvec{y} .
\end{array}
$$
Thus $X={\cal P} \circ \restric{Y}{M}=(1-ax)\restric{\tanvec{x}}{M}$
is the solution of the generalized nonholonomic system.

Let us study the infinitesimal symmetries of both systems.
We can see that a vector field $V \in {\cal X}(N)$ is an
infinitesimal symmetry of the unconstrained system if
it has the form \mbox{$V = V^1(ye^{-x})\tanvec{x}
+e^xV^2(ye^{-x})\tanvec{y}$}, where $V^1$ and
$V^2$ are arbitrary smooth functions.

On the other hand, since
the constrained system is one-dimensional,
its infinitesimal symmetries are the vector fields
$U=kX$, with $k \in \Real$. Observe that, in principle, an
infinitesimal symmetry of $Y$ does not lead to an
infinitesimal symmetry of $X$ by restriction to~$M$,
even when $\restric{Y}{M} \in \Vect(M)$.
Nevertheless, if we also require that
$V^T(C) \subset \Tan C$,
then we obtain $V^1(t)=k(1+a\ln(t/a))$ and $V^2(t)=0$,
so that actually
$\restric{V}{M}=k(1-ax)\restric{\tanvec{x}}{M}$
is an infinitesimal symmetry of $X$.

\subsection*{Example 2}

Here we discuss an example
of a particle with a nonholonomic constraint,
due to Rosenberg \cite{Ros-andy}.
Consider a particle moving
in $\Real^3$ with lagrangian function
$$
L=\frac{1}{2}(\dot{x}^2+\dot{y}^2+\dot{z}^2)
$$
subject to the nonholonomic constraint
$$
\phi=\dot{z}-y\dot{x}.
$$
Using the notation of section 5, we have
$N=\Tan\Real^3$,
$\omega_L=\dif x \wedge \dif \dot{x}+
\dif y \wedge \dif \dot{y}+
\dif z \wedge \dif \dot{z}$ and
$\dif E_L=\dot{x}\dif\dot{x}+
\dot{y}\dif\dot{y}+\dot{z}\dif\dot{z}$,
so the unconstrained dynamics is
the well-known free dynamics
described by the vector field
$$
X_L=\widehat\omega^{-1}_L(\dif E_L)
=\dot{x}\tanvec{x}+\dot{y}\tanvec{y}
+\dot{z}\tanvec{z}.
$$
The constraint submanifold is 
$M = \{ \dot{z}{=}y\dot{x} \}$,
with tangent bundle
$$
\Tan M=\Ker(\dif \phi)=
\restric{\left\langle\tanvec{x},\tanvec{y}+\dot{x}\tanvec{\dot{z}},
\tanvec{z},\tanvec{\dot{x}}+y\tanvec{\dot{z}},
\tanvec{\dot{y}}\right\rangle}{M},
$$
and the vector subbundle $C\subset\restric{\Tan N}{M}$ is
$$
C=\langle\widehat\omega^{-1}_L(\transp{J}(\dif \phi))\rangle =
\restric{\left\langle y\tanvec{\dot{x}}-\tanvec{\dot{z}}\right\rangle}{M} .
$$
Note that $\restric{\Tan N}{M}$ splits as
$\restric{\Tan N}{M}=\Tan M \oplus C$,
so the only solution $X$ of the constrained lagrangian
system is the projection of $\restric{X_L}{M}$ to $\Tan M$
according to this decomposition:
$$
X=\restric{ \left(\dot{x}\tanvec{x}+
\dot{y}\tanvec{y}+\dot{z}\tanvec{z}-
\frac{y\dot{y}\dot{x}}{y^2+1}\tanvec{\dot{x}}+
\frac{\dot{y}\dot{x}}{y^2+1}\tanvec{\dot{z}}\right)}{M}.
$$

We choose $(x,y,z,\dot{x},\dot{y})$ as coordinates on~$M$.
With this system, the vector field $X$ reads as
$$
X=\dot{x}\tanvec{x}+
\dot{y}\tanvec{y}+y\dot{x}\tanvec{z}-
\frac{y\dot{y}\dot{x}}{y^2+1}\tanvec{\dot{x}}.
$$
After some calculus, we can find the symmetries and
constants of motion of both systems. The constants
of motion of the free particle are the
functions $G(\dot{x},\dot{y},\dot{z},
\dot{x}y-\dot{y}x,\dot{y}z-\dot{z}y)$, where
$G$ is an arbitrary function with five variables.
The infinitesimal symmetries are linear combinations
of the six vector fields $\tanvec{x}$, $\tanvec{y}$,
$\tanvec{z}$, $x\tanvec{x}+\dot{x}\tanvec{\dot{x}}$,
$y\tanvec{y}+\dot{y}\tanvec{\dot{y}}$ and
$z\tanvec{z}+\dot{z}\tanvec{\dot{z}}$, with
the constants of motion as coefficients.

The constants of motion of the constrained system,
written in coordinates of~$M$, are
\beq
F\left( \dot{y},\dot{x}\sqrt{y^2+1},
\dot{y}x-\mathrm{arcsinh}(y)\dot{x}\sqrt{y^2+1},
\dot{y}z-\dot{x}(y^2+1) \right),
\label{constantF}
\eeq
and the infinitesimal symmetries are linear combinations
of the five vector fields
$$
\tanvec{x},\tanvec{z},
\dot{x}\tanvec{x}+\dot{y}\tanvec{y}+
y\dot{x}\tanvec{z}-
\frac{\dot{x}\dot{y}y}{y^2+1}\tanvec{\dot{x}},
$$
$$
\frac{\mathrm{arg\,sinh}(y)}{\dot{y}}\tanvec{x}+
\frac{\sqrt{y^2+1}}{\dot{y}}\tanvec{z}+
\frac{1}{\sqrt{y^2+1}}\tanvec{\dot{x}},
$$
$$
\frac{\dot{x}(y-\mathrm{arg\,sinh}(y)\sqrt{y^2+1})}{\dot{y}^2}\tanvec{x}+
\frac{y}{\dot{y}}\tanvec{y}-\frac{\dot{x}}{\dot{y}^2}\tanvec{z}-
\frac{\dot{x}y^2}{\dot{y}(y^2+1)}\tanvec{\dot{x}}+\tanvec{\dot{y}},
$$
with the constants of motion as coefficients.

\medskip

In order to illustrate proposition \ref{prop-constants}
we take a function $g=G(\dot{x},\dot{y},\dot{z},
\dot{x}y-\dot{y}x,\dot{y}z-\dot{z}y)$, i.e., a constant
of motion of~$X_L$, such that $Z \cdot g=0$, where $Z$ is
the section of $C$
$$
Z=\restric{X_L}{M}-X=
\restric{\frac{\dot{x}\dot{y}}{y^2+1}
\left( y\tanvec{\dot{x}}-\tanvec{\dot{z}} \right)}{M}.
$$
This yields to
$$
g=H\left(\dot{y},\sqrt{\dot{z}^2+\dot{x}^2},
\dot{z}+\dot{y}x-\dot{x}y-
\mathrm{arg\,sinh}(\dot{z}/\dot{x})\sqrt{\dot{z}^2+\dot{x}^2},
\dot{y}z-\dot{z}y-\dot{x}\right)
$$
and we see that $\restric{g}{M}$ is just the expression
(\ref{constantF}).

\appendix
\section*{Appendix: some lemmas about linear algebra}
\setcounter{section}{1}
\setcounter{equation}{0}
\def\thesection{\Alph{section}}

Here we collect some results about linear algebra
on vector bundles
that are needed in section~3.
These lemmas are stated and proved for vector spaces,
but of course nothing changes essentially
if vector bundles are considered instead.
$$
\xymatrix{
E \ar[r]^{f}  &  F \ar@{>>}[d]^{p} \\
E_o \ar@{^{(}->}[u]^{j} \ar[r]_{\bar f} & F\rlap{$/F_o$}
}
$$
\begin{lem}
Let $f \colon E \to F$ be a linear map between vector spaces,
and
$E_\circ \subset E$ and $F_\circ \subset F$ vector subspaces.
Denote $j \colon E_\circ \to E$ the inclusion,
$p \colon F \to F/F_\circ$ the projection to the quotient,
and consider the composition
$\bar f = p \circ f \circ j$.
Then:
\ben
\itemsep 0pt plus 1pt
\item
$\bar f$ is injective iff $E_\circ \cap f^{-1}(F_\circ) = \{0\}$.
\\
Assuming $f$ injective, this also amounts to
$f(E_\circ) \cap F_\circ = \{0\}$.
\item
$\bar f$ is surjective iff $f(E_\circ)+F_\circ = F$.
\\
Assuming $f$ surjective, this also amounts to
$E_\circ + f^{-1}(F_\circ) = E$.
\item
When $f$ is surjective,
$\bar f$ is bijective iff
$E_\circ \oplus f^{-1}(F_\circ) = E$.
\\
When $f$ is injective,
$\bar f$ is bijective iff
$f(E_\circ) \oplus F_\circ = F$.
\een
\end{lem}

\proof
First note that
\beq
\Ker \bar f = E_\circ \cap f^{-1}(F_\circ) , \quad
\Img \bar f = \big( f(E_\circ)+F_\circ \big) / F_\circ .
\eeq
These equalities are clear:
the kernel is constituted by the vectors in $E_\circ$ mapped to~$F_\circ$
by~$f$,
and the image of a subspace $F' \subset F$ by $p$ is
$(F'+F_\circ)/F_\circ$.
This readily yields the first assertions about injectivity and surjectivity.

Their equivalent formulations when $f$ is injective [or surjective]
can be proved using the formulas for
$f(E_1 \cap E_2)$ and $f^{-1}(F_1 \cap F_2)$
[or for the sum],
as well as
$f^{-1}(f(E_\circ))=E_\circ+\Ker f$,
$f(f^{-1}(F_\circ))=F_\circ \cap \Img f$.

Finally, the assertions about the bijectivity of~$\bar f$
are a trivial consequence of the other ones.
\qed

\medskip

Remember that a linear equation $f(x) = b$ is consistent iff $b \in \Img f$.
Now let us study a linear equation on~$E_\circ$
defined as in the preceding lemma
by $\bar f$ and the class of an element $b \in F$.
\begin{lem}
The linear equation
$\bar f(x) = \bar b$
is equivalent to the couple of equations
$f(x) - b \in F_\circ$, $x \in E_\circ$.
It is consistent iff $b \in f(E_\circ)+F_\circ$;
in this case the solution is unique iff
$E_\circ \cap f^{-1}(F_\circ) = \{0\}$.
\qed
\end{lem}

\medskip

Finally, let $E \subset G$ be a subspace of a vector space.
Recall that the \textit{annihilator} (or orthogonal) of $E$ is the subspace
$$
E^\vdash =
\{ \gamma \in G^* \mid (\forall x \in E)\; \langle \gamma,x \rangle = 0 \}
\subset G^* .
$$
This space has a close relationship with $G/E$.
Indeed,
the transpose map of $G \to G/E$ defines a canonical isomorphism
$$
\delta \colon (G/E)^* \to E^\vdash ,
$$
such that, for $\alpha \in E^\vdash$ and $z \in G$,
$\langle \delta^{-1}(\alpha) , z+E \rangle =
\langle \alpha , z \rangle$.

\begin{lem}
Let $E, F \subset G$ be vector subspaces.
Let $(\alpha^1,\ldots,\alpha^p)$
be a basis for the annihilator $E^\vdash \subset G^*$,
and $(v_1,\ldots,v_q)$
a basis for~$F$.
Consider the matrix
$D = (D^i_{\,j})_{{1 \leq i \leq p \atop 1 \leq j \leq q}}$
with elements
$D^i_{\,j} = \langle \alpha^i, v_j \rangle$.
Then:
\ben
\itemsep 0pt plus 1pt
\item
$E + F = G$ iff $\rank D = p$.
\item
$E \cap F = \{0\}$ iff $\rank D = q$.
\item
$E \oplus F = G$ iff $D$ is square invertible.
\een
\end{lem}

\proof 
Consider the linear map $\eps \colon F \to G/E$ defined as
the composition of the inclusion $F \inmap{} G$ and the projection
to the quotient $G \surmap{} G/E$. It is clear that $E + F = G$
iff $\eps$ is surjective, and $E \cap F = \{0\}$ iff $\eps$ is
injective, so the only thing to prove is that the given matrix is
the matrix $D$ of~$\eps$ in appropriate bases: the basis $(v_j)$
for~$F$, and the basis $(\bar\alpha_i)$, the dual basis of
$\bar\alpha^i = \delta^{-1}(\alpha^i)$, for $G/E$.

Then, if
$\eps(v_j) = \bar\alpha_i \, D^i_{\,j}$,
we have
$
D^i_{\,j} =
\langle \bar\alpha^i, \eps(v_j) \rangle =
\langle \alpha^i, v_j \rangle
$,
which is what we wanted to prove.
\qed

\subsection*{Acknowledegments}

The authors acknowledge partial financial support from 
project BFM2002--03493.



\begin{thebibliography}{ABCD99}
\small
\itemsep 0pt plus 1pt

\bibitem[Arn\,83]{Arn-dynsyst3}
{\sc V.\,I. Arnol'd (ed.)},
{\sl Dynamical Systems III},
Encyclop\ae dia of Math.\ Sciences~3,
Springer, Berlin, 1988.

\bibitem[BKMM\,96]{BKMM-sym}
{\sc A.\,M. Bloch, P.\,S. Krishnaprasad, J.\,E. Marsden, R.\,M. Murray},
``Nonholonomic mechanical systems with symmetry'',
{\sl Arch. Rational Mech. Anal. \bf 36} (1996) 21--99.

\bibitem[B\'S\,93]{BS-nonh}
{\sc L. Bates, J. \'Sniatycki},
``Nonholonomic reduction'',
{\sl Rep. Math. Phys. \bf 32} (1993) 99--115.

\bibitem[BGPR\,86]{BGPR-equiv}
{\sc C. Batlle, J. Gomis, J.\,M. Pons and N. Rom\'an-Roy},
``Equivalence between the lagrangian and hamiltonian formalisms
for constrained systems'',
{\sl J.~Math. Phys. \bf 27} (1986) 2953--2962.

\bibitem[Car\,90]{Car-theory}
{\sc J.\,F. Cari\~nena},
``Theory of singular lagrangians'',
{\sl Fortschrit. Phys.~\bf 38} (1990) 641--679.

\bibitem[CR\,93]{CR-multipliers}
{\sc J.\,F. Cari\~nena and M.\,F. Ra\~nada},
``Lagrangian systems with constraints: a geometric approach to
the method of Lagrange multipliers''.
{\sl J. Phys. A: Math. Gen. \bf 26} (1993) 1335--1351.

\bibitem[CL\,99]{CL-redrec}
{\sc J. Cort\'es, M. de Le\'on},
``Reduction and reconstruction of the dynamics of
nonholonomic systems'',
{\sl J.~Phys. A: Math. Gen. \bf 32} (1999) 8615--8645.

\bibitem[GNH\,78]{GNH-pres}
{\sc M.\,J. Gotay, J.\,M. Nester and G. Hinds},
``Presymplectic manifolds and the Dirac-Bergmann theory of constraints'',
{\sl J.~Math. Phys. \bf 19} (1978) 2388--2399.

\bibitem[GP\,89]{GP-K}
{\sc X. Gr\`acia and J.\,M. Pons},
``On an evolution operator connecting lagrangian and hamiltonian formalisms'',
{\sl Lett. Math. Phys. \bf 17} (1989) 175--180.

\bibitem[GP\,91]{GP-unif}
{\sc X. Gr\`acia and J.\,M. Pons},
``Constrained systems: a unified geometric approach'',
{\sl Int. J. Theor. Phys. \bf 30} (1991) 511--516.

\bibitem[GP\,92]{GP-gener}
{\sc X. Gr\`acia and J.\,M. Pons},
``A generalized geometric framework for constrained systems'',
{\sl Diff. Geom. Appl.~\bf 2} (1992) 223--247.

\bibitem[GP\,01]{GP-struc}
{\sc X. Gr\`acia and J.\,M. Pons},
``Singular lagrangians: some geometric structures along the Legendre map'',
{\sl J.~Phys.~A: Math. Gen. \bf 34} (2001) 3047--3070.

\bibitem[GP\,02]{GP-sym}
{\sc X. Gr\`acia and J.\,M. Pons},
``Symmetries and infinitesimal symmetries of singular differential equations'',
{\sl J.~Phys.~A: Math. Gen. \bf 35} (2002) 5059--5077.

\bibitem[GPR\,91]{GPR-higher}
{\sc X. Gr\`acia, J.\,M. Pons and N. Rom\'an-Roy},
``Higher order lagrangian systems:
geometric structures, dynamics, and constraints'',
{\sl J.~Math. Phys. \bf 32} (1991) 2744--2763.

\bibitem[KM\,01]{KM-rel}
{\sc O. Krupkov\'a and J. Musilov\'a},
``The relativistic particle as a mechanical system with non-holonomic
constraints'',
{\sl J.~Phys.~A: Math. Gen. \bf 34} (2001) 3859--3875.

\bibitem[ILMM\,96]{ILMM-implicit}
{\sc L.\,A. Ibort, M. de Le\'on, G. Marmo amd D. Mart\'{\i}n de Diego},
``Non-holonomic constrained systems as implicit differential equations'',
{\sl Rend. Sem. Mat. Univ. Politec. Torino \bf 54} (1996) 295--317.

\bibitem[Koi\,92]{Koi-red}
{\sc J. Koiller},
``Reduction of some classical non-holonomic systems with symmetry'',
{\sl Arch. Rational Mech. Anal. \bf 118} (1992) 113--48.

\bibitem[KM\,98]{KM-Poisson}
{\sc W.\,S. Koon and J.\,E. Marsden},
``Poisson reduction for nonholonomic mechanical systems with symmetry'',
{\sl Rep. Math. Phys. \bf 42} (1998) 101--134.

\bibitem[LM\,96]{LM-lag}
{\sc M. de Le\'on, D. Mart\'in de Diego},
``On the geometry of non-holonomic lagrangian systems'',
{\sl J.~Math. Phys. \bf 37} (1996) 3389--414.

\bibitem[Mar\,98]{Mar-approaches}
{\sc C.-M. Marle}
``Various approaches to conservative and nonconservative nonholonomic systems'',
{\sl Rep. Math. Phys. \bf 42} (1998) 211--229.

\bibitem[Mar\,03]{Mar-sym}
{\sc C.-M. Marle}
``On symmetries and constants of motion in hamiltonian systems with nonholonomic constraints'',
{\sl Classical and quantum integrability (Warsaw, 2001)}, 223--242,
Banach Center Publ., 59, 
Polish Acad. Sci., Warsaw, 2003.

\bibitem[MVB\,02]{MVB-nonh}
{\sc E. Massa, S. Vignolo and D. Bruno},
``Non-holonomic lagrangian and hamiltonian mechanics:
an intrinsic approach'',
{\sc J.~Phys.~A: Math. Gen. \bf 35} (2002) 6713--6742.

\bibitem[MMT\,95]{MMT-integrability}
{\sc G. Mendella, G. Marmo and W.\,M. Tulczyjew},
``Integrability of implicit differential equations'',
{\sl J.~Phys.~A: Math. Gen. \bf 28} (1995) 149--163.

\bibitem[MT\,78]{MT-ham}
{\sc M.\,R. Menzio and W.\,M. Tulczyjew},
``Infinitesimal symplectic relations and
generalized hamiltonian dynamics'',
{\sl Ann. Inst. Henri Poincar\'e~A \bf 28} (1978) 349--367.

\bibitem[Ros\,77]{Ros-andy}
{\sc R. Rosenberg},
{\sl Analytical Dynamics},
Plenum, New York, 1977.

\bibitem[Tul\,86]{Tul-constraints}
{\sc W.\,M. Tulczyjew},
``Differential geometry of mechanical systems with constraints'',
{\sl Atti Accad. Sci. Torino Cl. Sci. Fis. Mat. Natur. \bf 120}
(1986) 211--216.

\bibitem[VF\,72]{VF-diff}
{\sc A.\,M. Vershik, L.\,D. Faddeev},
``Differential geometry and lagrangian mechanics with constraints'',
{\sl Sov. Phys. Dokl. \bf 17} (1972) 34--36.



\end{thebibliography}
\end{document}